\documentclass[%
 reprint,
superscriptaddress,
 amsmath,amssymb,
 aps,
]{revtex4-2}

\usepackage{microtype}
\usepackage{amsmath}
\usepackage{graphicx}
\usepackage{dcolumn}
\usepackage{bm}
\usepackage{adjustbox} 
\usepackage{graphicx}

\usepackage{color, hyperref}
\definecolor{mygrey}{gray}{0.80}
\definecolor{darkblue}{RGB}{8,81,156}
\hypersetup{colorlinks,breaklinks,
            linkcolor=darkblue,urlcolor=darkblue,
            anchorcolor=darkblue,citecolor=darkblue}

\definecolor{super-dark-green}{RGB}{0,69,41}
\definecolor{super-dark-purple}{RGB}{63,0,125}
\definecolor{super-dark-blue}{RGB}{8,48,107}
\definecolor{super-dark-red}{RGB}{165,0,38}

\setcitestyle{super}

\begin{document}


\title{The Impact of Hydration Shell Inclusion and Chain Exclusion in the Efficacy of Reaction
Coordinates for Homogeneous and Heterogeneous Ice Nucleation}

\author{Kimia Sinaeian}
\affiliation{%
 Department of Chemical and Environmental Engineering, Yale University, New Haven, Connecticut 06520, United States
}%

\author{Amir Haji-Akbari}
\email{Corresponding author. Email: amir.hajiakbaribalou@yale.edu}
\affiliation{%
 Department of Chemical and Environmental Engineering, Yale University, New Haven, Connecticut 06520, United States
}%

\date{\today}

\begin{abstract}
\noindent
Ice nucleation plays a pivotal role in many natural and industrial processes, and molecular simulations play have proven vital in uncovering its kinetics and mechanisms. A fundamental component of such simulations is the choice of an order parameter (OP) that quantifies the progress of nucleation, with the efficacy of an OP typically measured by its ability to predict the committor probabilities. Here, we leverage a machine learning framework introduced in our earlier work (Domingues,~\emph{et al.}, \emph{J. Phys. Chem. Lett.}, 15, 1279, {\bf 2024}) to systematically investigate how key implementation details influence the efficacy of standard Steinhardt OPs in capturing the progress of both homogeneous and heterogeneous ice nucleation. Our analysis identify distance and $q_6$ cutoffs, as the primary determinants of OP performance, regardless of the mode of nucleation. We also examine the impact of two popular refinement strategies, namely chain exclusion and hydration shell inclusion, on OP efficacy. We find neither strategy to exhibit a universally consistent impact. Instead, their efficacy depends strongly on the chosen distance and $q_6$ cutoffs. Chain exclusion enhances OP efficacy when the underlying OP lacks sufficient selectivity, whereas hydration shell inclusion is beneficial for overly selective OPs. Consequently, we demonstrate that selecting optimal combinations of such  cutoffs can eliminate the need for these refinement strategies altogether. These findings provide a systematic understanding of how to design and optimize OPs for accurately describing complex nucleation phenomena, offering valuable guidance for improving the predictive power of molecular simulations.
\end{abstract}

\maketitle


    \section{Introduction}
    \label{section:intro}

\noindent
    Crystallization is a phenomenon that is key to many scientifically and technologically important processes, from cloud microphysics and precipitation in the atmosphere\cite{Baker1997, Atkinson2013, Cziczo2013, Storelvmo2017} to the production of nanoparticles,\cite{BassaniACSNano2024} pharmaceuticals,\cite{Chen2011} semiconductors,\cite{LeeNatRevMater2016} solar cells\cite{UdayabhaskararaoChemMater2017, SharmaNanoscaleAdv2021} and aviation equipment.\cite{Potapczuk2013}  As a first-order phase transition, crystallization typically proceeds through a nucleation and growth mechanism.\cite{DeYoreo2003} Nucleation is the initial-- and often the rate-limiting-- step wherein a sufficiently large crystalline nucleus-- known as critical nucleus-- emerges within the liquid, and its kinetics and mechanism impact the properties of the arising crystals, such as grain size and polymorph composition. Despite its significance, there are considerable gaps in our understanding of the microscopic mechanisms of nucleation.\cite{VekilovCrystGrowthDes2010, Sosso2016} This is primarily due to the activated nature of nucleation, which becomes a rare event\cite{Hussain2020} when nucleation barriers exceed $kT$ considerably. As such, conventional experimental techniques lack sufficient resolution to probe the mechanism of localized swift structural rearrangements that underlie nucleation. Moreover, due to long wait times preceding nucleation, traditional molecular simulation techniques such as molecular dynamics\cite{Alder1959} (MD) and Monte Carlo\cite{Metropolis1953} (MC) are inefficient in capturing improbable fluctuations that result in nucleation. These considerations pose major challenges to experimental and computational studies of crystal nucleation.

Over recent decades, advanced sampling techniques, such as umbrella sampling,\cite{TorrieJComputPhys1977} metadynamics,\cite{LaioPNAS2002} transition interface sampling\cite{vanErpJCP2003} (TIS) and forward flux sampling\cite{AllenJCP2005} (FFS) have become indispensable for elucidating the free energetics, kinetics, and mechanisms of homogeneous and heterogeneous crystal nucleation.\cite{VanDuijneveldtJCP1992, ReintenWolde1996, SalvalaglioPNAS2015, HajiAkbari2015, BonatiPRL2018, ArjunJPCB2020, HallArxiv2024} Among these methods, FFS stands out for its ability to directly calculate nucleation rates from unbiased trajectories, even in driven systems. FFS has been extensively employed to investigate crystal nucleation across a diverse range of systems,  such as water,\cite{Li2011, HajiAkbari2014, Cabriolu2015,  HajiAkbari2015, Sosso2016hetIN, Bi2016, HajiAkbari2017, LiNatComm2017, HussainJACS2021, HussainJCP2021} silicon,\cite{LiJCP2009, LiNatMater2009} salts,\cite{ValerianiJCP2005, JiangJCP2018, JiangJCP2018p} Lennard-Jones,\cite{MithenJCP2014, HussainJCP2022} hard spheres,\cite{FilionJCP2010, RichardJCP2018} and beyond.\cite{BiJPCB2014} All such calculations rely on the availability of an order parameter (OP)-- also known as a collective variable (CV)-- that quantifies the progress of nucleation. Their accuracy and efficiency therefore rely on the extent to which the chosen OPs also function as effective reaction coordinates (RCs),~i.e.,~whether they accurately predict the committor probability, which represents the likelihood that a trajectory initiated from a given configuration will reach the target state.\cite{PetersAnnRevPhysChem2016}  The most effective CVs are typically those that are physically interpretable and that closely align with the committor probability. Beyond improving the accuracy and efficiency of rare event simulations, well-designed OPs also hold promise for experimental studies, particularly in colloidal self-assembly, where real-time tracking of particle trajectories is feasible.\cite{Gasser2001, Savage2009}

In crystal nucleation studies, the size of the largest crystalline nucleus is typically used as the OP of choice.\cite{LupiJPCL2017} Crystalline nuclei are commonly identified using Steinhardt bond order parameters\cite{Steinhardt1983} (BOPs), which effectively distinguish the structural arrangements of bulk fluid particles from those within crystalline phases. The identification process, however, is complex and typically involves selecting multiple cutoffs and applying additional refinement strategies, guided by a mix of empiricism and physical intuition.  In our recent work,\cite{Domingues2024} we introduced a systematic methodology that leverages high-throughput screening and machine learning to evaluate the sensitivity of OP performance to variations in cutoffs and implementation details. This approach diverges from earlier methods,\cite{PetersJCP2007} which utilized statistical tools such as maximum likelihood estimators to construct optimal combinations of pre-defined OPs. Instead, our methodology focuses on statistically rigorous feature selection and optimization to enhance OP efficacy. While the primary objective of Ref.~\citenum{Domingues2024} was to refine structural OPs that exhibit limitations in describing the heterogeneous nucleation of close-packed crystals, we demonstrate the broader utility of this approach for systematically identifying and optimizing features critical to nucleation studies.

In this work, we employ the framework introduced in Ref.~\citenum{Domingues2024} to systematically evaluate standard OPs for ice nucleation, a process of critical importance in diverse fields such as atmospheric physics\cite{Murray2012} and cryopreservation.\cite{HajiAkbariPNAS2016, You2020, Lin2023} We consider both homogeneous nucleation in the bulk and heterogeneous nucleation on a graphene surface. In addition to assessing the standard cutoffs central to the implementation of the BOP-based OPs, we investigate the efficacy of two widely adopted refinement strategies: chain exclusion\cite{ReinhardtJCP2012} and hydration shell inclusion.\cite{Li2011} Our analysis reveals a striking similarity in the behavior of OPs between homogeneous and heterogeneous nucleation scenarios. Furthermore, we find that neither refinement strategy consistently improves or diminishes OP efficacy. Instead, their performance appears to heavily depend on the selectivity of the unrefined OPs.

This paper is organized as follows. The employed methodology, including the details of MD simulations, OP construction, screening and assessment, is outlined in Section~\ref{section:methods}. Our findings are presented in Section~\ref{section:results} while Section~\ref{section:conclusion} summarizes our key conclusions as well as making some broad observations. 

    \section{Methods}
    \label{section:methods}
    
\subsection{System Description and Molecular Dynamics Simulations}
\label{section:methods:system}

\noindent
We consider homogeneous and heterogeneous ice nucleation within the coarse-grained monoatomic water\cite{Molinero2008} (mW) system, with heterogeneous nucleation considered on a single layer of graphene with interaction parameters given by Bi~\emph{et al}.\cite{Bi2016} All MD  simulations are conducted using  \texttt{LAMMPS}\cite{Plimpton1995, Thompson2022} and equations of state are integrated using the velocity Verlet algorithm with a time step of 5~fs. In the case of homogeneous nucleation, simulations are conducted in the isothermal isobaric (NPT) ensemble at 225~K and 1~atm, with temperature and pressure controlled using the Nos\'{e}-Hoover thermostat\cite{Nose1983ConstantSystems} and the Parrinello-Rahman barostat,\cite{Parrinello1981} respectively. In the case of heterogeneous nucleation, simulations are conducted in the canonical (NVT) ensemble at 240~K wherein a supported liquid film is exposed to a rigid graphene layer. As explained earlier,\cite{HussainJACS2021} this corresponds to a simulation at an effective zero pressure as the free interface can adjust freely to accommodate density fluctuations.

\subsection{Order Parameters for Ice Nucleation}
    
\begin{figure}
\centering
\includegraphics[width=.43\textwidth]{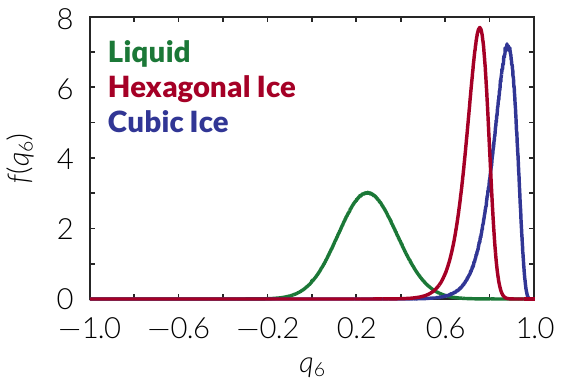}
\caption{\label{fig:q6}$q_6$ profiles of water molecules within the bulk liquid, and cubic and hexagonal ice obtained at 225~K and 1~bar.}
\end{figure}

\noindent
An order parameter, or collective variable, is a mechanical observable $\lambda: \mathcal{Q} \rightarrow \mathbb{R}(\mathbb{Z})$ that maps each configuration $x \in \mathcal{Q}$ onto a real (or an integer) number, $\lambda(x)$, that quantifies  the progress of a rare event. 
Similar to the procedure employed in Ref.~\citenum{Domingues2024}, we utilize the committor probabilities calculated in Section \ref{section:methods:system} to systematically evaluate the sensitivity of ice nucleation  OPs to various implementation details. As noted in Section~\ref{section:intro}, the size of the largest crystalline nucleus is the most commonly employed OP to track the progress of crystal nucleation.  This requires the definition of an appropriate local measure of crystallinity, typically based on Steinhardt BOPs.\cite{Steinhardt1983}  Starting from $\{\mathbf{r}_i\}_{i=1}^{n}$, the positions of all building blocks within a given configuration, the nearest neighbors of each building block are identified using a proximity measure. Most commonly, particles within a distance $r_c$ from the $i$-th particle are labeled as its nearest neighbors, where $r_c$ is usually selected as the position of the first minimum in the radial distribution function of the supercooled liquid (or supersaturated solution). Subsequently, a set of complex-valued vectors is computed as follows:
\begin{eqnarray}
q_{lm}(i) &=& \frac{1}{N_b(i)} \sum_{j=1}^{N_b(i)} Y_{lm}(\theta_{ij},\phi_{ij}),~~ -l\le m\le l.
\end{eqnarray}
Here, $N_b(i)$ is the number of nearest neighbors of the $i$-th particle, and $\theta_{ij}$ and $\phi_{ij}$ are the polar and azimuthal angles associated with the vector connecting $i$ to its $j$th nearest neighbor. $Y_{lm}(\cdot)$'s are spherical harmonic functions given by: 
\begin{eqnarray}
Y_{lm}(\theta,\phi) &=& (-1)^m\sqrt{
\frac{2l+1}{4\pi}\frac{(l-m)!}{(l+m)!}
}P_{lm}(\cos\theta)e^{im\phi}
\end{eqnarray}
where $P_{lm}$'s are the associated Legendre polynomials obtained by,
\begin{eqnarray}
P_{lm}(x) &=& \frac{(-1)^m\left(1-x^2\right)^{\frac m2}}{2^ll!}\frac{d^{l+m}}{dx^{l+m}}\left(x^2-1\right)^{l}.
\end{eqnarray}
Conceptually, the vector, $\mathbf{q}_l(i)=\left(q_{l,-l}(i),\cdots,q_{l,l}(i)\right)$ encodes the orientational signature of the neighbors surrounding the central particle $i$. Once computed for all building blocks, these $\mathbf{q}_l$ vectors can be further refined through additional algebraic operations, such as neighbor-averaging\cite{LechnerJCP2008} or coarse-graining.\cite{Domingues2024} Nonetheless, the arising $\mathbb{C}^{2l+1}$ vectors will still vary upon rotating the nearest neighbor shell around a central particle. Therefore, it is necessary to transform them into \emph{scalar invariants} that remain unchanged under rigid-body rotations. Since there are multiple distinct ways of constructing such invariants, it is important to identify a suitable set of invariants that unambiguously distinguish the supercooled (or supersaturated) state from all plausible crystals. For nucleation of ice and other tetrahedral crystals, this is achieved by computing the mean normalized  inner-product of $\mathbf{q}_6$ across the first nearest neighbor shell:
\begin{eqnarray}
q_6(i) &=& \frac{1}{N_b(i)}\sum_{j=1}^{N_b(i)}\frac{\mathbf{q}_6(i)\cdot\mathbf{q}_6^*(j)}{\left|\mathbf{q}_6(i)\right|\left|\mathbf{q}_6(j)\right|}
\end{eqnarray}
As illustrated in Fig.~\ref{fig:q6}, the histograms of $q_6(i)$ values within the supercooled liquid are clearly separated from those corresponding to hexagonal and cubic ice. This separation makes it possible to define a threshold, $q_{6,c}$, such that any molecule with $q_6(i)\ge q_{6,c}$ is classified as 'solid-like`. Subsequently, solid-like particles that are within a distance $r_{c,c}(\le r_c)$ of each other are clustered together into graphs referred to as  \emph{crystalline nuclei}. The size of the largest nucleus, determined by the number of nodes (particles) it contains,  is then utilized as the OP to monitor the progress of nucleation.

It is important to note that this procedure is not unique. Beyond its sensitivity to implementation details-- such as the choice of $l$, the scalar invariant(s), $r_c$, $q_c$ and $r_{c,c}$-- alternative methods for defining local crystallinity have been proposed, including BOPs constructed from real-valued spherical harmonic functions,\cite{ReinhardtJCP2012p} and OPs that do not rely on spherical harmonics altogether, such as alignment-based nematic OPs.\cite{Yi2013} For ice nucleation, these BOPs are further augmented with additional refinement strategies. Here, we focus on two such strategies. The first strategy, \emph{chain exclusion}, was proposed by Reinhardt~\emph{et al.}\cite{ReinhardtJCP2012} and addresses the formation of long-lived ice-like chains in supercooled water. In this method, such ice-like chains are pruned from the nucleus in order to make it more compact. More specifically, within nuclei larger than 10 water molecules, ice-like molecules with only one ice-like neighbor are removed unless their sole neighbor has at least three ice-like neighbors. This process is applied recursively until no such molecules remain in the nucleus. The second strategy, \emph{hydration shell inclusion},\cite{Li2011}  compensates for the limitations of bulk-based OPs in identifying interfacial molecules. In this approach, liquid-like molecules that are first-nearest neighbors of ice-like molecules within a cluster are also included as part of the cluster.

In summary, this elaborate procedure entails the specification of five cutoffs and decision points, the distance cutoff ($r_c$), the $q_6$ cutoff ($q_{6,c}$), the clustering distance cutoff ($r_{c,c}$), and the application (or omission) of chain exclusion ($c$) and hydration shell inclusion ($h$). These five features collectively impact the process of constructing crystalline nuclei and, consequently, the size of the largest crystalline nucleus,  $\lambda(x;r_c,q_{6,c},r_{c,c},c,h)$.

\subsection{Order Parameter Assessment}
\label{section:methods:op-assessment}

\noindent
In order to analyze the sensitivity of OP efficacy to the above-mentioned features, we generate 110,000 different OPs, each corresponding to a unique combination of feature values. We assess each OP based on its ability to predict the committor probability, $p_c(x)$, which is the probability that a trajectory initiated from a particular configuration $x\in\mathcal{Q}$ reaches the target crystalline basin (before returning to the liquid). More precisely, committor probabilities on level sets of an effective OP are expected to be narrowly distributed. Such an analysis is, however, only meaningful if conducted over a collection of configurations with a diverse range of $p_c$'s, spanning the full range between zero and unity. In order to generate such a collection, we start from the crossing configurations obtained during our prior calculations of homogeneous\cite{HajiAkbari2018} and heterogeneous\cite{HussainJACS2021} ice nucleation rates for the mW\cite{Molinero2008} system using the jumpy forward flux sampling\cite{HajiAkbari2018} (jFFS) algorithm. More specifically, we initiate 500-ps MD trajectories from a subset of configurations with $\approx50\%$ survival probability\cite{Domingues2024} saving configurations every 2.5 ps. Such trajectories end up descending towards either of the supercooled liquid and crystalline basins, yielding configurations with a diverse range of committor probabilities. We randomly select $n_c=1,000$ of the arising configurations and directly compute their committor probabilities by launching $N_t=50$ momenta-randomized trajectories from each configuration and enumerating the fraction of those that first reach the crystalline basin.  In this analysis, we use the OP employed in the original rate calculation and define the 'crystalline basin` in a conservative manner to ensure that our $p_c$ estimates are not affected by the potential lack of efficacy of the utilized OP. 

We then conduct high-throughput screening of the feature space by computing the largest cluster sizes for the same set of configurations using the 110,000 unique combinations of feature values. We assess each OP by the $R^2$ of the following nonlinear regression problem:
\begin{eqnarray}
\min_{a,b}\sum_{i=1}^{n_c}w_i\left\{
\frac12\Big\{
1+\text{erf}\,\left[
a(\lambda(x_i)-b)
\right]
\Big\}-p_c(x_i)
\right\}^2,\label{minimization_problem} 
\end{eqnarray}
Here, $n_c$ is the number of configurations, $p_c(x_i)$ and $w_i$ are the committor probability and weight of the $i$-th configurations, respectively, and $a$ and $b$ are fitting parameters. In other words, Eq.~\eqref{minimization_problem} quantifies the mean squared error (MSE) between actual committors, and those predicted using an error functional model,\cite{Ma2005} which is a reasonable representation of $p_c$ for a rare event with a single dominant free energy barrier.\cite{Wedekind2007,  Peters2010}

\begin{figure}
\centering
\includegraphics[width=.39\textwidth]{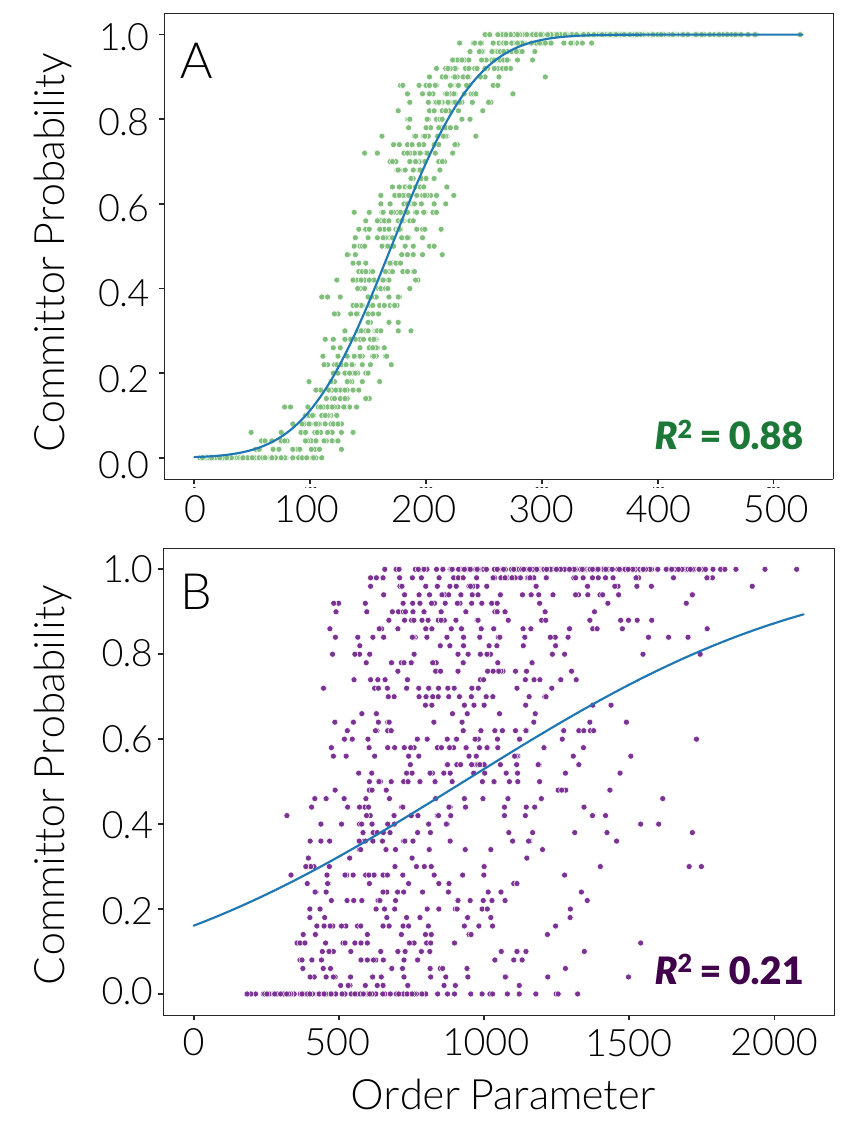}
\caption{\label{fig.1} Assessing OP efficacy based on committor probability distributions. For an effective OP (A), configurations with the same OP value exhibit narrow $p_c$ distributions, enabling accurate  $p_c$ predictions using Eq.~(\ref{minimization_problem}). In contrast, a poor OP (B) results in broad $p_c$ distributions, leading to a poor fit with a small $R^2$.}
\end{figure}

The weights for individual configurations are assigned based on their committor probabilities, grouping them into "decades``. For instance, configurations with $0\le p_c<0.1$ belong to the first decade, and so forth.  The weight of each configuration is inversely proportional to the number of configurations in its respective decade. This approach ensures that configurations with small or large committor probabilities-- overrepresented in our sample-- do not artificially inflate the $R^2$  value of the MSE regression. Such configurations are typically easy to distinguish, even with suboptimal OPs, as they are far from the transition region. By applying weighted regression, we aim to more accurately resolve the efficacy of the OPs in capturing the progress of nucleation (near the transition state).

Figure~\ref{fig.1} depicts a scatterplot of $p_c(x_i)$ versus $\lambda(x_i)$, comparing the effectiveness of two different OPs. In Fig.~\ref{fig.1}A, the chosen OP is effective: $p_c$'s of configurations with identical $\lambda$ values exhibit a narrow distribution, indicating that the OP is highly predictive of $p_c$.  Additionally, an error-functional committor model accurately captures this relationship, achieving an $R^2$ value of 0.88. In contrast, the OP depicted in Fig.~\ref{fig.1}B is very ineffective, as configurations with identical $\lambda$ values exhibit a broad $p_c$ distribution. Accordingly, the committor model poorly describes the data in this case, reflected by a low $R^2$ value of 0.21.

After computing $R^2$ values for all OPs, we apply machine learning (ML) techniques to evaluate the relative importance of various features.\cite{LundbergArXiv2017} We first train an Extreme Gradient Boosting\cite{chen2016xgboost} (XGBoost) model from the Python library \texttt{scikit-learn}\cite{PedregosaJMachLearnRes2011} (Version 1.1.3) using 80\% of the dataset, reserving the remaining 20\% for testing to assess the model’s generalizability. XGBoost is a robust ML algorithm that iteratively builds decision trees to correct previous errors, enhancing predictive accuracy. Unlike the random forest\cite{breiman2001random} approach, which independently constructs decision trees and average their predictions, XGBoost builds trees sequentially to minimize errors in each step. It further optimizes traditional gradient boosting by incorporating regularization to reduce overfitting and speeding up computation. To ensure the reproducibility of the model results, the \texttt{random\_state} parameter is set to $0$. To prevent any node in the training set from having too few points, the maximum depth and the number of boosted trees are set to $10$ and $100$, respectively.  Repeating our analysis using a random forest model,\cite{breiman2001random} yields comparable results. To quantify each feature's significance, we calculate SHapley Additive exPlanations (SHAP) scores,\cite{lundberg2017unified} which provide a game-theoretic breakdown of feature contributions to the objective function-- in this case $R^2$.  Specifically, the SHAP score for a given feature reflects the expected change in $R^2$ when that feature’s value is specified.  An advantage of SHAP scores is their capacity to capture each feature's impact on the objective function for a particular combination of other features' values, accounting for correlations among features. Features that yield consistently large SHAP scores (positive or negative) are more likely to influence the objective function in a statistically significant manner. Accordingly, we employ the absolute mean SHAP score as a qualitative measure of a feature’s importance in determining OP efficacy.

\begin{figure*}
\centering
\includegraphics[width=.99\textwidth]{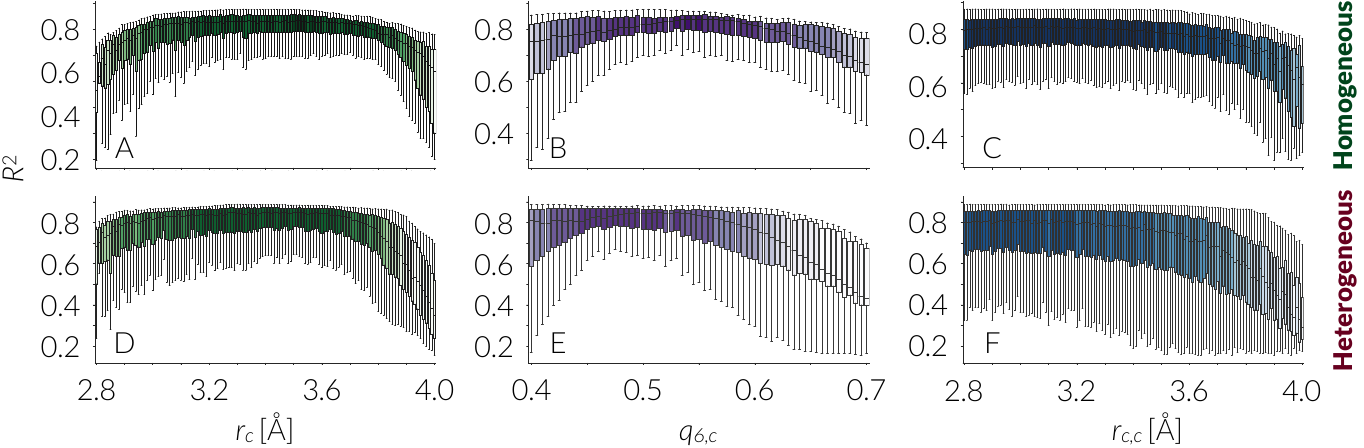}
\caption{\label{fig3:cutoffs} Box plots of the $R^2$'s of distinct OPs as a function of (A,D) distance, (B,E) $q_6$, and (C,F) clustering distance cutoffs for (A-C) homogeneous and (D-F) heterogeneous nucleation.  }
\end{figure*}

    \section{Results and Discussion}
    \label{section:results}

\subsection{Effect of Individual Features}

\begin{figure*}
\centering
\includegraphics[width=.98\textwidth]{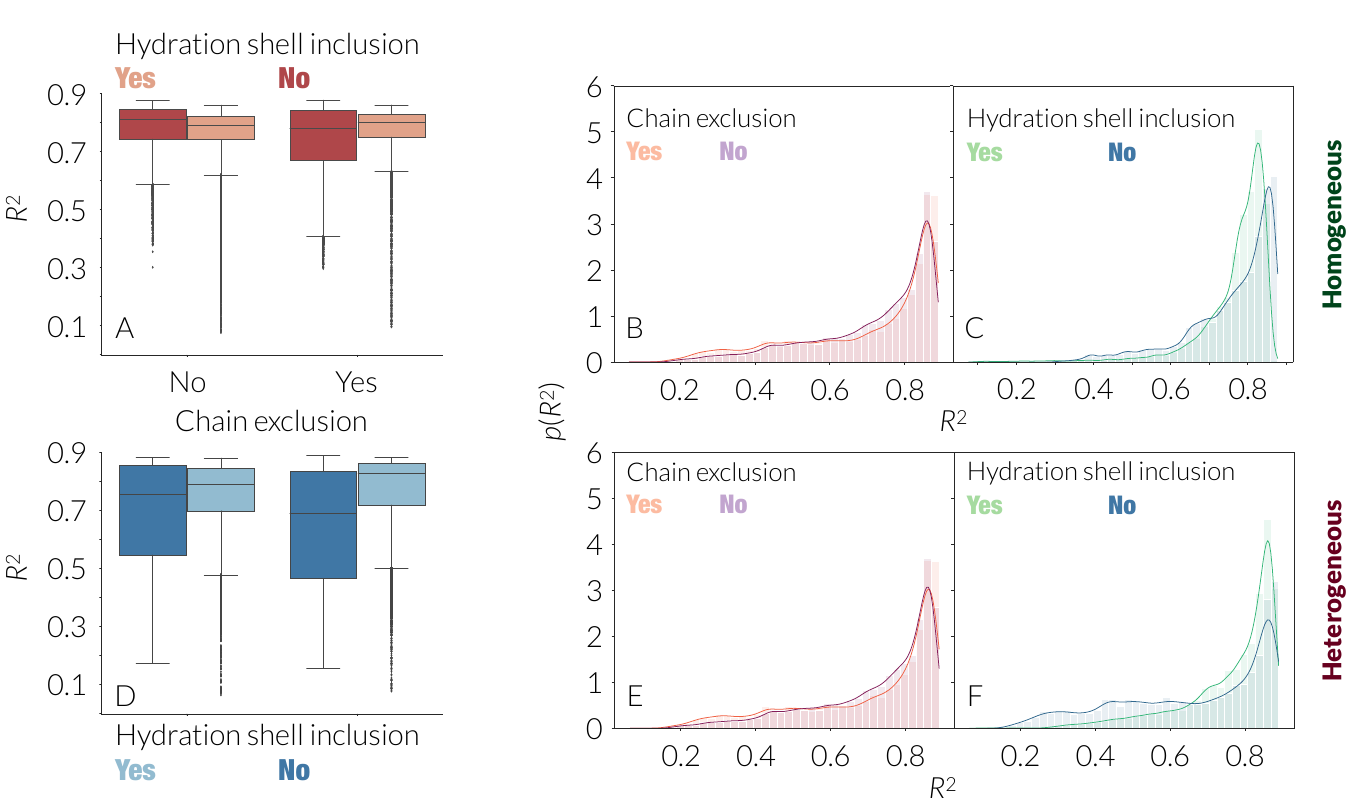}
\caption{\label{fig4:CE-HSI}(A-B) Box plots of $R^2$ for the collective effect of chain exclusion and hydration shell inclusion on (A) homogeneous and (B) heterogeneous nucleation. (C-F) $R^2$ probability density functions in the presence and absence of (C,E) chain exclusion and (D,F) hydration shell inclusion in (C-D) homogeneous and (E-F) heterogeneous nucleation.}
\end{figure*}

\noindent
We first examine how OP efficacy varies with individual feature values. Figures~\ref{fig3:cutoffs}A-F depict box plots of the average $\langle R^2 \rangle$ as a function of  $r_c$, $q_{6,c}$, and $r_{c,c}$ for both homogeneous and heterogeneous nucleation. Distance cutoffs within the 3.1–3.7~\AA~range produce effective OPs for both homogeneous (Fig.~\ref{fig3:cutoffs}A) and heterogeneous (Fig.~\ref{fig3:cutoffs}D) nucleation. This is expected, as this range aligns with the first valley of the radial distribution function. Outside this range, OP efficacy declines and becomes more sensitive to changes in other feature values, leading to increased variability. Specifically, reducing $r_c$ below 3.1~\AA~can lead to missed nearest neighbors and an inaccurate characterization of the orientational structure surrounding a central molecule. 
When it comes to clustering distance cutoffs (Figs.~\ref{fig3:cutoffs}C,F), smaller $r_{c,c}$ values generally yield better OPs, manifest in larger $R^2$'s. What is notable though is the relatively larger variability in $R^2$ due to the fact that each $r_{c,c}$ value encompasses all $r_c$'s that are larger, henceforth including OPs with a wide range of distance cutoffs. In other words, $r_c$ and $r_{c,c}$ are highly correlated, making any inference about the impact of $r_{c,c}$ on $R^2$ nontrivial.  When it comes to $q_{6,c}$, the most effective OPs are attained with $0.5\le q_{6,c}\le 0.6$ for homogeneous nucleation (Fig.~\ref{fig3:cutoffs}B) and $0.45\le q_{6,c}\le 0.55$ for heterogeneous nucleation (Fig.~\ref{fig3:cutoffs}E). These values are consistent with the separation between the liquid and crystalline $q_6$  histograms depicted in Fig.~\ref{fig:q6}. Using $q_{6,c}$ values outside this range compromises the selectivity of the order parameter, resulting in the mislabeling of liquid-like or solid-like particles.

\begin{figure*}
\centering
\includegraphics[width=.85\textwidth]{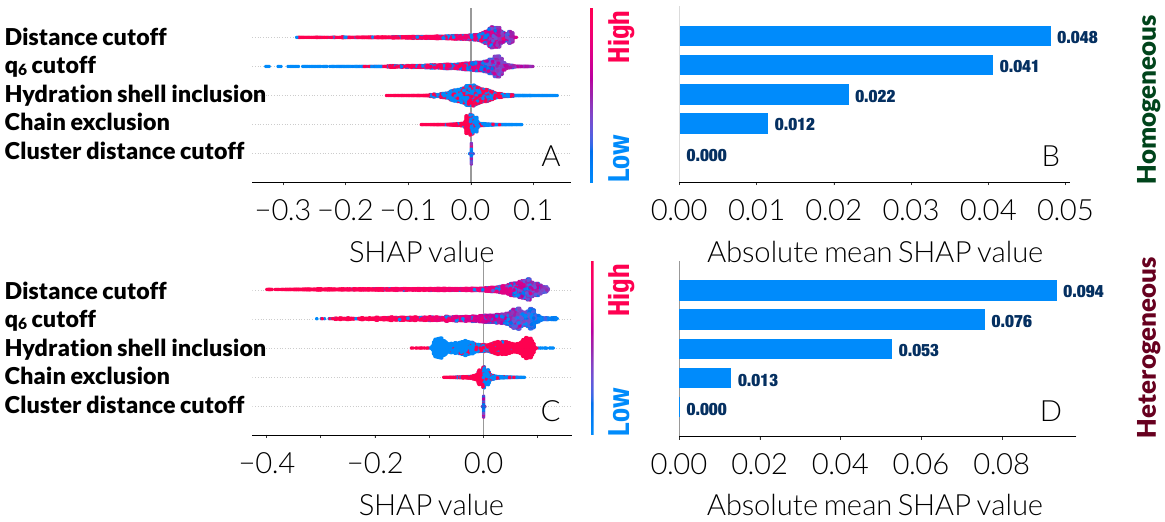}
\caption{\label{fig5:SHAP}(A,C) Swarm plots and (B, D) absolute mean SHAP values of the five features needed for constructing OPs for (A-B) homogeneous and (C-D) heterogeneous nucleation.}
\end{figure*}

It must also be noted that these optimal cutoffs are independent of the mode of nucleation, and are almost identical for homogeneous and heterogeneous nucleation. This is in contrast to simpler liquids\cite{Domingues2024} (such as the LJ and hard sphere systems) and might be attributed to the tetrahedral structure of water that is not as strongly perturbed by the graphene surface. This is also consistent with the overwhelming body of prior work that generally demonstrates that OPs specifically designed for studying homogeneous ice nucleation are effective at describing the physics of heterogeneous nucleation.\cite{LupiJCP2016, LupiJPCL2017}

As depicted in Fig.~\ref{fig4:CE-HSI}, applying chain exclusion (Figs.~\ref{fig4:CE-HSI}B,E) or hydration shell inclusion (Figs.~\ref{fig4:CE-HSI}B,E) does not alter the $R^2$ distributions considerably for neither homogeneous (Figs.~\ref{fig4:CE-HSI}B-C) nor heterogeneous (Figs.~\ref{fig4:CE-HSI}E-F) nucleation. As such, neither strategy appears to be uniformly effective in improving OP efficacy. Moreover, the simultaneous application of these strategies (or lack thereof) appears ineffective in improving OP efficacy, as evident from the box plots depicted in Figs.~\ref{fig4:CE-HSI}A,D. These box plots denote the data quartiles, as well as outliers that fall outside the second and third quartiles. If anything, applying hydration shell inclusion (irrespective of whether chain exclusion is applied or not) leads to more outliers. In other words, applying hydration shell inclusion makes the OP more sensitive to other implementation details, resulting on some remarkably poor OPs. 

\subsection{Feature Selection via Machine Learning}

\noindent
These observations provide preliminary valuable insights into the individual contributions of different features to OP efficacy; however, they do not rigorously address potential correlations among these features. To overcome this limitation, we employ machine learning to compute SHAP value distributions for various features, following the methodology described in Section~\ref{section:methods:op-assessment}, with the results presented in Fig.~\ref{fig5:SHAP}. Not surprisingly, $r_c$ and $q_{6,c}$ emerge as the most influential features in determining OP efficacy. Specifically, changing the values of these features leads to pronounced changes in SHAP scores, as illustrated by the swarm plots of Figs.~\ref{fig5:SHAP}A,C. Furthermore, the absolute mean SHAP values depicted in Figs.~\ref{fig5:SHAP}B,D reveal that $r_c$ and $q_{6,c}$ account for 39\% and 33\% of the $R^2$ variations in both homogeneous and heterogeneous nucleation, respectively. Notably, the greater importance of $r_c$ relative to the $q_{6,c}$  contrasts with findings for nucleation of close-packed crystals, where the BOP cutoff was at least twice as significant as the distance cutoff.\cite{Domingues2024} This discrepancy likely stems from the tetrahedral structure of water, where selecting an $r_c$ that accurately captures the first hydration shell of a central molecule inherently identifies neighbors that are already tetrahedrally ordered. Such tetrahedral arrangement within the first hydration shell aligns well with the long-range order characteristic of cubic and hexagonal ice. In contrast, for systems forming close-packed crystals, the particle arrangements within the first coordination shell often deviate significantly from the target crystal structure at the local level, making the distance-based identification of order less effective.

\begin{figure}
\centering
\includegraphics[width=.5\textwidth]{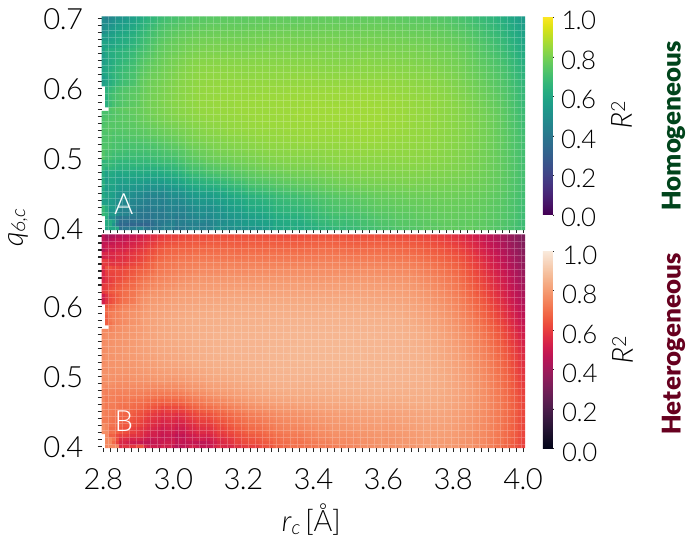}
\caption{\label{fig6:correlation}Heat maps of mean $R^2$ as a function of $r_c$ and $q_{6,c}$, the two most important features per SHAP analysis for (A) homogeneous and (B) heterogeneous nucleation.}
\end{figure}

\begin{figure*}
\centering
\includegraphics[width=.9\textwidth]{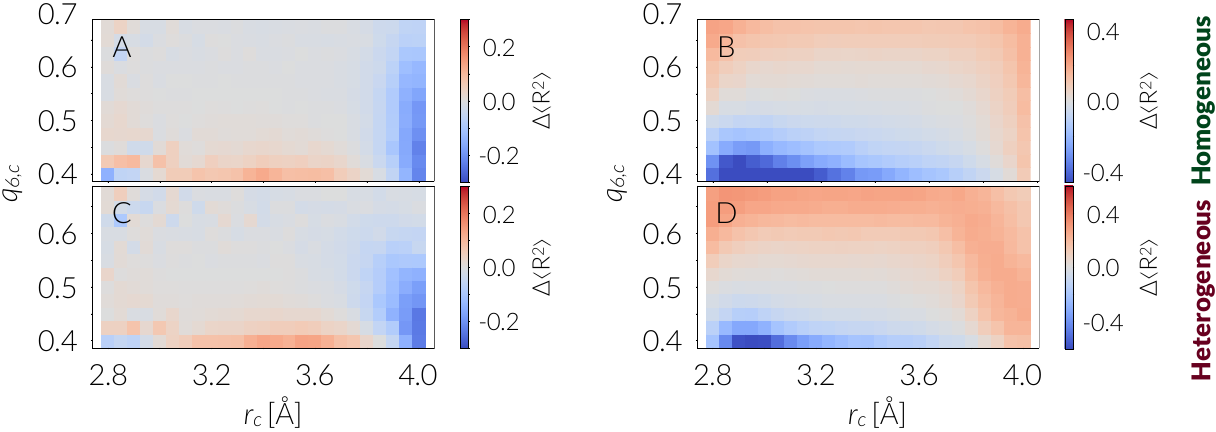}
\caption{\label{fig7:correlation-ce-hsi}Heat maps of changes in mean $R^2(r_c,q_{6,c})$ in the presence and in the absence of (A,C) chain exclusion and (B,D) hydration shell inclusion for (A-B) homogenous and (C-D) heterogeneous nucleation.}
\end{figure*}

Both chain exclusion and hydration shell inclusion exhibit substantial absolute mean SHAP scores, respectively accounting for 18\% and 10\% of $R^2$ variability in homogeneous nucleation, and 22\% and 6\% in heterogeneous nucleation. This observation is intriguing, given that neither strategy appears to have a significant global impact on OP efficacy as shown in Fig.~\ref{fig4:CE-HSI}. These findings suggest that the influence of each strategy on OP efficacy depends on the values of other features and varies across the feature space rather than being uniformly distributed. This highlights the strength of SHAP analysis in uncovering complex feature correlations that might remain obscured when the objective function is projected onto a single feature dimension.

Among the five features analyzed here, $r_{c,c}$ is the only one that does not exhibit a noticeable impact on OP efficacy. This finding is consistent with our previous conclusion for close-packed crystals,\cite{Domingues2024} which indicated that choosing a different distance cutoff for clustering does not lead to significant improvements in OP efficacy. Thus, it is reasonable to infer that this strategy may not be worthwhile, except in cases involving crystals where the building blocks are separated by distances that are vastly different from the characteristic size of the first coordination shell within the metastable fluid.

\begin{figure}
\centering
\includegraphics[width=.37\textwidth]{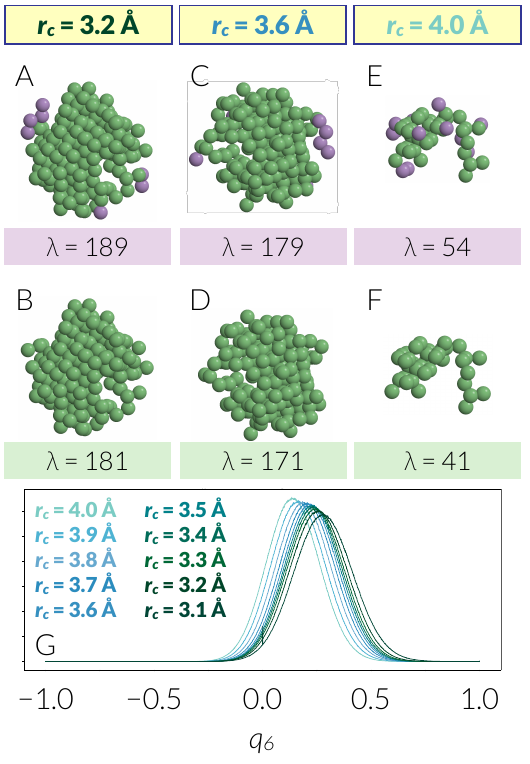}
\caption{\label{fig8:example-CE}(A-F) An illustration of the combined effect of distance cutoff and chain exclusion on the crystalline nucleus identified using a distance cutoff of (A-B) 3.2~\AA, (C-D) 3.6~\AA and (E-F) 4.0~\AA~(A,C,E) without and (B,D,F) with chain exclusion for a representative configuration obtained for homogeneous nucleation. Water molecules removed as chains are shown in light purple in (A,C,E).  (G) The sensitivity of $q_6$ histograms within the bulk liquid to distance cutoff, indicating a leftward shift upon increasing $r_c$.}
\end{figure}

To further investigate the intricacies of correlations within the feature space, we compute the mean $R^2$ as a function of $r_c$ and $q_{6,c}$, identified by SHAP analysis as the two most influential features. Figure~\ref{fig6:correlation} presents heat maps of $R^2(r_c, q_{6,c})$ for homogeneous  (Fig.~\ref{fig6:correlation}A) and heterogeneous  (Fig.~\ref{fig6:correlation}B) nucleation. In both cases, the optimality band-- i.e.,~the region of the $(r_c,q_{6,c})$ space that produces the most effective OPs-- appears curved and tilted, indicating correlations between $r_c$ and $q_{6,c}$. This observation is consistent with previous findings for LJ and hard sphere systems,\cite{Domingues2024} underscoring the necessity of optimizing distance and BOP cutoffs concurrently rather than estimating each feature's optimal value independently. Furthermore, the heat maps exhibit qualitative similarity and do not appear to be sensitive to the mode of nucleation, suggesting that the graphene surface considered here does not considerably alter the shape and the locus of the optimality band. 

We then explore how the impact of chain exclusion and hydration shell inclusion on OP efficacy depends on $r_c$ and $q_{6,c}$. Figure~\ref{fig7:correlation-ce-hsi} depicts heat maps of $\Delta{R}^2$, the change in mean $R^2$ upon applying the corresponding strategy. Red regions signify improved OP efficacy in the presence of the said strategy, while blue regions indicate deterioration. We observe that chain exclusion is most effective when the underlying OP is not selective enough,~e.g.,~at lower values of $q_{6,c}$ that produce OPs lacking sufficient selectivity, for both homogeneous (Fig.~\ref{fig7:correlation-ce-hsi}A) and heterogeneous (Fig.~\ref{fig7:correlation-ce-hsi}C) nucleation. (Here, poor selectivity refers to the inclusion of molecules with insufficiently crystalline local environments within the nucleus.) In these cases, chain exclusion compensates for the poor selectivity of the underlying OP. Conversely, the biggest deterioration in OP efficacy occurs when the OP is overly selective, such as at excessively high $q_{6,c}$ values. This is consistent with the main premise of chain exclusion,~i.e.,~the removal of molecules with a local tetrahedral arrangement participating in chains of spuriously crystalline molecules.

In contrast, hydration shell inclusion exhibits the opposite behavior. By adding the first hydration shell of the original nucleus to the nucleus, hydration shell inclusion improves the performance of overly selective OPs but exacerbates the shortcomings of OPs that lack sufficient selectivity (Figs.~\ref{fig7:correlation-ce-hsi}B,D). This behavior is intuitive, as adding additional molecules to a nucleus that already contains an excess of spuriously solid-like molecules is expected to diminish OP efficacy. It must be noted that these observations are independent of the mode of nucleation, and hold true for both homogeneous  (Figs.~\ref{fig7:correlation-ce-hsi}A-B) and heterogeneous (Figs.~\ref{fig7:correlation-ce-hsi}C-D) nucleation.

\begin{figure}
\centering
\includegraphics[width=.37\textwidth]{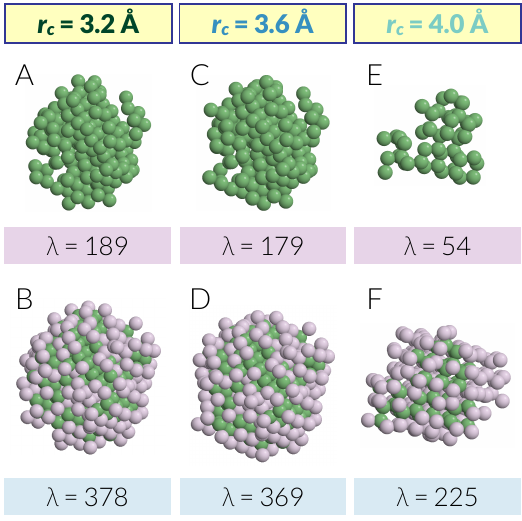}
\caption{\label{fig9:example-HSI}(A-F) An illustration of the combined effect of distance cutoff and hydration shell inclusion on the crystalline nucleus identified using a distance cutoff of (A-B) 3.2~\AA, (C-D) 3.6~\AA~and (E-F) 4.0~\AA~(A,C,E) without and (B,D,F) with hydration shell inclusion for a representative configuration obtained for homogeneous nucleation. Water molecules that belong to the first hydration shell of (A,C,E) are depicted in light purple in (B,D,F). }
\end{figure}

We wish to point out the connection between these refinement strategies and pruning,\cite{Domingues2024} another refinement approach that we had previously proposed and assessed for the heterogeneous nucleation of close-packed crystals. Notably, our prior findings demonstrated that pruning is most effective when the underlying OP lacks sufficient selectivity. Collectively, these observations suggest that the effectiveness of a refinement strategy depends on the selectivity of the original OP and the nature of refinement-- specifically, whether it involves adding to or removing building blocks from the nucleus identified without the strategy. Strategies that enlarge the nucleus are more effective when the OP is overly selective, while those that shrink the nucleus are better suited for OPs that lack sufficient selectivity. From a practical perspective, it is more effective to focus on optimizing the original OP directly (by choosing the right combination of distance and BOP cutoffs) rather than relying on innovative or complex refinement strategies. Refinement strategies should be considered primarily as backup options, employed only when the development of a straightforward, optimized OP proves infeasible, such as heterogeneous nucleation on surfaces that induce considerable lateral ordering within the fluid.

A peculiar observation in Fig.~\ref{fig7:correlation-ce-hsi} is the apparent over-selectivity of OPs that employ distance cutoffs larger than $\approx3.8\,$\AA, as they improve upon hydration shell inclusion but deteriorate upon chain exclusion. Indeed, applying an unrefined OP with a distance cutoff of 4\,\AA~to a typical configuration results in significantly smaller nuclei (Figs.~\ref{fig8:example-CE}C and \ref{fig9:example-HSI}C) compared to those obtained with smaller $r_c$  values (Figs.~\ref{fig8:example-CE}A-B and \ref{fig9:example-HSI}A-B). Unsurprisingly, chain exclusion further reduces the nucleus size (Fig.~\ref{fig8:example-CE}F), while hydration shell inclusion partially restores some of the excluded molecules (Fig.~\ref{fig9:example-HSI}F).

These results can be attributed to the fact that using an $r_c$  beyond the first valley of the radial distribution function includes molecules from the second hydration shell among nearest neighbors. This inclusion disrupts the coherence of the tetrahedral order between the central molecule and its first nearest neighbors. Consequently, the local $q_6$ histograms of molecules within the supercooled liquid shift leftward as $r_c$ increases, as illustrated in Fig.~\ref{fig8:example-CE}G. For sufficiently large crystalline nuclei, this loss of coherence has minimal impact on their cores, where the first few hydration shells of constituent molecules remain crystalline, maintaining a clear separation between the first and second hydration shells. However, towards the surfaces of the nuclei, the disrupted coherence misclassifies solid-like molecules as liquid-like, leading to an underestimation of nucleus size. Therefore, the application of chain exclusion further removes putative chains from the nucleus core, while hydration shell inclusion recovers some of the outermost molecules misclassified due to the over-selectivity of the OP.

\section{Conclusions}
\label{section:conclusion}

\noindent
In conclusion, we use MD simulations, forward-flux sampling, committor analysis, and machine learning to systematically evaluate the performance of OPs for computational studies of ice nucleation. We screen over $10^5$ OPs, each utilizing a unique combination of cutoffs and refinement strategies to identify the size of the largest crystalline nucleus and assess their efficacy in predicting committor probabilities for both homogeneous and heterogeneous ice nucleation in the mW system.  In addition to standard distance and BOP cutoffs essential to implementing traditional Steinhardt OPs, we investigate the impact of two widely used refinement strategies: exclusion of tetrahedral chains from the crystalline nucleus and inclusion of the first hydration shell within the nucleus. Through Shapley value analysis, we identify standard distance and $q_6$ cutoffs as the most influential features, accounting for  72\% of the variability in $R^2$
 for committor probability predictions. Although chain exclusion and hydration shell inclusion do not appear to exhibit a global impact on OP efficacy, they explain around 28\% of the variability according to our SHAP analysis. A closer examination of the sensitivity of OP efficacy on distance and $q_6$ cutoffs reveals that chain exclusion enhances OP performance when the unrefined OP lacks sufficient selectivity but reduces efficacy when the OP is overly selective.  Conversely, hydration shell inclusion improves over-selective OPs while further degrading those with insufficient selectivity. These findings underscore that these refinement strategies do not possess any inherent value but can be instrumental in scenarios where the unrefined OP lacks the appropriate level of selectivity. 

It is important to highlight that there are alternative methodologies for constructing OPs for ice nucleation, ranging from traditional approaches based on Steinhardt BOPs\cite{MoorePCCP2010, NguyenJPCB2015} to more sophisticated data science-based techniques.\cite{GeigerJCP2013, DeFeverChemSci2019, TakahashiPCCP2023} All these methods rely on specific cutoffs and implementation details, typically determined through a combination of physical intuition and trial-and-error. The systematic screening approach presented in this study offers a robust framework for evaluating the relative importance of such cutoffs and implementation details, and determining their optimal values, thereby enhancing the reliability and efficacy of OP construction across methodologies.

It is also critical to acknowledge the potential limitations of our CV assessment strategy. First of all, the employed error-functional committor model of Eq.~\eqref{minimization_problem} assumes that ice nucleation constitutes a single-step transition over a sufficiently large parabolic barrier, with the system evolving diffusively along the reaction coordinate.\cite{Wedekind2007, Peters2010} There are alternative committor models in the literature, such as the logistic function.\cite{Peters2006} Moreover, instead of the MSE approach utilized here, one can instead adopt a generalized maximum likelihood framework for parametric optimization.\cite{Peters2006, Peters2010} It is possible that no simplified committor model may fully capture the complexity of nucleation, which can involve multistep transitions, multidimensional reaction coordinates, or asymmetric pathways. Additionally, the choice of optimization approach may influence the quantitative ranking of OPs.  Nevertheless, prior studies on crystal nucleation within melts suggest that it is often a single-step process satisfactorily described by an error functional model.\cite{Beckham2011, LupiJCP2016, LupiNature2017, LupiJPCL2017} As such, the inferred OP efficacy is fairly insensitive to  the specific committor model or optimization method employed.\cite{Domingues2024} While these limitations might slightly alter the computed scores, we do not expect them to affect the key conclusions of this work.

A key observation of this study is the striking similarity in OP efficacy behavior between homogeneous and heterogeneous nucleation. This is consistent with previous investigations confirming the transferability of standard OPs, originally developed for homogeneous ice nucleation, to heterogeneous nucleation across various surfaces.\cite{LupiJCP2016, LupiJPCL2017} However, recent studies\cite{ZhaoJCP2023} have shown that excessive interfacial ordering can significantly impair OP efficacy. Extending the systematic feature selection approach employed here to evaluate the effectiveness of various refinement strategies for constructing robust OPs for heterogeneous nucleation on such surfaces presents a valuable avenue for future research.

\begin{acknowledgments}
\noindent
A.H.-A. gratefully acknowledges the support of National Science Foundation (NSF) Grants CBET-1751971 (CAREER Award) and CHE-2203527. These calculations were performed at the Yale Center for Research Computing (YCRC).
\end{acknowledgments}

\bibliographystyle{apsrev} 

\bibliography{References-Revised}

\end{document}